\RequirePackage{ifpdf}
\ifpdf 
\documentclass[pdftex]{sigma}
\else
\documentclass{sigma}
\fi

\begin{document}

\allowdisplaybreaks

\renewcommand{\thefootnote}{$\star$}

\renewcommand{\PaperNumber}{075}

\FirstPageHeading

\ShortArticleName{On Initial Data in the Problem of Consistency on
Cubic Lattices for $3 \times 3$ Determinants}

\ArticleName{On Initial Data in the Problem of Consistency\\ on
Cubic Lattices for $\boldsymbol{3 \times 3}$ Determinants\footnote{This
paper is a contribution to the Proceedings of the Conference ``Symmetries and Integrability of Dif\/ference Equations (SIDE-9)'' (June 14--18, 2010, Varna, Bulgaria). The full collection is available at \href{http://www.emis.de/journals/SIGMA/SIDE-9.html}{http://www.emis.de/journals/SIGMA/SIDE-9.html}}}

\Author{Oleg I.~MOKHOV~$^{\dag\ddag}$}

\AuthorNameForHeading{O.I.~Mokhov}

\Address{$^\dag$~Centre for Nonlinear Studies, L.D.Landau Institute for Theoretical Physics,\\
\hphantom{$^\dag$}~Russian Academy of Sciences, 2 Kosygina Str., Moscow, Russia}
\EmailD{\href{mailto:mokhov@mi.ras.ru}{mokhov@mi.ras.ru}, \href{mailto:mokhov@landau.ac.ru}{mokhov@landau.ac.ru}, \href{mailto:mokhov@bk.ru}{mokhov@bk.ru}}

\Address{$^\ddag$~Department of Geometry and Topology, Faculty of Mechanics and Mathematics,\\
\hphantom{$^\ddag$}~M.V.~Lomonosov Moscow State University, Moscow, Russia}

\ArticleDates{Received January 23, 2011, in f\/inal form July 17, 2011;  Published online July 26, 2011}

\Abstract{The paper is devoted to complete proofs of theorems on consistency
on cubic lattices for $3 \times 3$ determinants. The discrete
nonlinear equations on $\mathbb{Z}^2$ def\/ined by the condition that
the determinants of all $3 \times 3$ matrices of values of the
scalar f\/ield at the points of the lattice $\mathbb{Z}^2$ that form
elementary $3 \times 3$ squares vanish are considered; some explicit
concrete conditions of general position on initial data are
formulated; and for arbitrary initial data satisfying these concrete
conditions of general position, theorems on consistency on cubic
lattices (a consistency ``around a cube'') for the considered discrete nonlinear equations on~$\mathbb{Z}^2$ def\/ined by $3 \times 3$ determinants are proved.}


\Classification{39A05; 52C07; 15A15; 37K10; 11H06}

\renewcommand{\thefootnote}{\arabic{footnote}}
\setcounter{footnote}{0}

\section{Introduction}

In this paper we present complete proofs of theorems on consistency
on cubic lattices for $3 \times 3$ determinants. Formulations and a
scheme of proofs of these theorems were given by the author in \cite{1,
2}, where a new, modif\/ied, consistency principle on cubic lattices
for a special class of two-dimensional discrete equations def\/ined by
relations on elementary $N \times N$ squares of the square lattice
$\mathbb{Z}^2$, $N > 2$, was proposed. Earlier, in \cite{3, 4, 5}, the
remarkable and very natural {\it principle of consistency on cubic
lattices} was proposed for discrete equations def\/ined by relations
on elementary $2 \times 2$ squares of the square lattice
$\mathbb{Z}^2$ as an ef\/fective test singling out a certain special
class of ``integrable'' discrete equations (see also \cite{6, 7, 8, 9, 10, 11, 12, 13, 14}). In
this paper, we consider only the discrete nonlinear equations on
$\mathbb{Z}^2$ def\/ined by the condition that the determinants of all
$3 \times 3$ matrices of values of the corresponding scalar f\/ield
$u$ at the points of the lattice $\mathbb{Z}^2$ that form elementary
$3 \times 3$ squares vanish. We formulate some explicit concrete
conditions of general position on initial data, and for arbitrary
initial data satisfying these concrete conditions of general
position, we prove theorems on consistency on cubic lattices (a consistency ``around a cube'') for the
considered discrete nonlinear equations on $\mathbb{Z}^2$ def\/ined by
$3 \times 3$ determinants.

\section{Consistency principle on cubic lattices}

We consider the square lattice $\mathbb{Z}^2$ consisting of all points
with arbitrary integer coordinates in $\mathbb{R}^2 = \{ (x_1, x_2)| \ x_k \in
\mathbb{R},\ k =1, 2 \}$ and complex (or real) scalar f\/ields $u$ on
the lattice $\mathbb{Z}^2$, $u : \mathbb{Z}^2 \rightarrow
\mathbb{C}$, def\/ined by their values $u_{i_1 i_2}$, $u_{i_1 i_2} \in
\mathbb{C}$, at each lattice point with coordinates $(i_1, i_2)$,
$i_k \in \mathbb{Z}$, $k = 1, 2$.

We consider a class of two-dimensional discrete equations on the
lattice $\mathbb{Z}^2$ for the f\/ield $u$ that are given by functions
$Q (x_1, x_2, x_3, x_4)$ of four variables with the help of the
relations
\begin{gather} Q (u_{i j}, u_{i + 1, j}, u_{i, j + 1}, u_{i + 1, j +
1}) = 0, \qquad i, j \in \mathbb{Z}, \label{1}
\end{gather}
so that in each
{\it elementary $2 \times 2$ square of the lattice $\mathbb{Z}^2$},
i.e., in each set of lattice points with coordinates of the form
$\{(i, j), (i + 1, j), (i, j + 1), (i + 1, j + 1)\}$, $i, j \in
\mathbb{Z}$, the value of the f\/ield $u$ at one of the vertices of
the square is determined by the values of the f\/ield at the other
three vertices.

In this case the scalar f\/ield $u$ on the lattice $\mathbb{Z}^2$ is
completely determined by f\/ixing initial data, for example, on the
coordinate axes of the lattice, $u_{i \, 0}$ and $u_{ 0 j}$, $i, j
\in \mathbb{Z}$.

Here, we do not discuss conditions on the initial data $u_{i j}$
themselves that must correctly and completely determine a scalar
f\/ield $u$ on the lattice $\mathbb{Z}^2$ for concrete discrete
equations of the form~\eqref{1}, and also we do not discuss
conditions on the functions $Q (x_1, x_2, x_3, x_4)$ for which
relations~\eqref{1} correctly determine a two-dimensional discrete
equation on the lattice $\mathbb{Z}^2$ for the f\/ield $u$.

We consider the cubic lattice $\mathbb{Z}^3$ consisting of points
with integer coordinates in $\mathbb{R}^3 = \{ (x_1, x_2, x_3)\,|\, x_k
\in \mathbb{R},\;  k =1, 2, 3 \}$ and f\/ix initial data, for example,
on the coordinate axes of the lattice, $u_{i\, 0 0}$, $u_{ 0 j\,
0}$, and $u_{ 0 0 k}$, $i, j, k \in \mathbb{Z}$.

A two-dimensional discrete equation~\eqref{1} is said to be {\it
consistent on the cubic lattice} (see \cite{3, 4, 5, 6, 7}) if for generic
initial data the discrete equation~\eqref{1} can be satisf\/ied in a
consistent way simultaneously on all two-dimensional {\it
coordinate} sublattices of the cubic lattice~$\mathbb{Z}^3$ that are
def\/ined by f\/ixing one of coordinates (any of the three coordinates)
of the cubic lattice. This condition is equivalent to the
consistency condition on each {\it elementary $2 \times 2 \times 2$
cube of the lattice~$\mathbb{Z}^3$}, $\{(i + p, j + r, k + s),  0
\leq p, r, s \leq 1\}$, where $i, j,$ and $k$ are arbitrary f\/ixed
integers, $i, j, k \in \mathbb{Z}$, i.e., relation~\eqref{1} must be
satisf\/ied in a consistent way on all faces of any elementary $2
\times 2 \times 2$ cube of the lattice~$\mathbb{Z}^3$ for generic
initial data. In the elementary cube $\{(i, j, k), 0 \leq i, j, k
\leq 1 \}$ the values $u_{1 0 1}$, $u_{1 1 0}$, and $u_{0 1 1}$ are
determined by relations~\eqref{1} on the corresponding faces of the
cube by the initial data $u_{0 0 0}$, $u_{1 0 0}$, $u_{0 1 0}$, and
$u_{0 0 1}$, and three relations on three faces of the cube must be
satisf\/ied for the value $u_{1 1 1}$. One can consider the condition
of consistency of the overdetermined system of relations for the
value~$u_{1 1 1}$ for generic initial data as the consistency
condition for the discrete equation~\eqref{1} on the cubic lattice
$\mathbb{Z}^3$.

Here, we do not discuss all various situations in which relations~\eqref{1}
correctly def\/ine a two-dimensional discrete equation for
the f\/ield $u$ on any two-dimensional coordinate sublattice of the
cubic lattice $\mathbb{Z}^3$; for example, one can assume for
simplicity that relations~\eqref{1} are invariant with respect to
the full symmetry group of the square. Classif\/ications of discrete
equations of the form~\eqref{1} that are consistent on the cubic
lattice were studied in \cite{6} and \cite{10} under some additional
conditions on the functions $Q (x_1, x_2, x_3, x_4)$ (see also
\cite{11}). The equation
\begin{gather} u_{i, j + 1} u_{i + 1, j} - u_{i + 1, j + 1}
u_{i j} = 0, \qquad i, j \in \mathbb{Z}, \label{2}
\end{gather}
def\/ined by
the condition that the determinants of all $2\times2$ matrices of
values of the f\/ield $u$ at the vertices of elementary $2\times2$
squares of the lattice $\mathbb Z^2$ vanish, is an example of such a~two-dimensional nonlinear discrete equation that is consistent on
the cubic lattice. Equation~\eqref{2} is linear with respect to each
variable and invariant with respect to the full symmetry group of
the square. Fixing arbitrary nonzero initial data $u_{i 0}$ and
$u_{0 j}$, $i, j \in \mathbb{Z}$, on the coordinate axes of the
lattice $\mathbb{Z}^2$ completely determines a f\/ield $u$ on the
lattice $\mathbb{Z}^2$ satisfying the discrete nonlinear equation~\eqref{2},
and f\/ixing arbitrary nonzero initial data $u_{i  0 0}$,
$u_{ 0 j 0}$, and $u_{0 0 k}$, $i, j, k \in \mathbb{Z}$, on the
coordinate axes of the lattice $\mathbb{Z}^3$ completely determines
a f\/ield $u$ on the lattice $\mathbb{Z}^3$ satisfying the discrete
nonlinear equation~\eqref{2} on all two-dimensional coordinate
sublattices of the cubic lattice~$\mathbb{Z}^3$; i.e., relations~\eqref{2}
are satisf\/ied in a consistent way on all faces of each
elementary $2 \times 2 \times 2$ cube of the lattice~$\mathbb{Z}^3$
for arbitrary nonzero initial data. The integrability (in the
broadest sense of the word) of the discrete nonlinear equation~\eqref{2}
is obvious, since it can be easily linearized: $ \ln u_{i,
j + 1} + \ln u_{i + 1, j} - \ln u_{i + 1, j + 1} - \ln u_{i j} = 0$,
$i, j \in \mathbb{Z}$.
Discrete nonlinear equations def\/ined by determinants of higher
orders are also $\mathcal{C}$-integrable and it is not a problem
to write their general solution for generic initial data, but in any case they are much more complicated, nonlinearizable,
and the problem on their consistency on cubic
lattice is very nontrivial (see~\cite{1, 2}).

\section[Relations on elementary $3 \times 3$ squares of the
lattice $\mathbb{Z}^2$ and consistency conditions]{Relations on elementary $\boldsymbol{3 \times 3}$ squares of the
lattice $\boldsymbol{\mathbb{Z}^2}$\\ and consistency conditions}

We will use the following def\/inition everywhere in this paper. An
{\it elementary $N \times N$ square of the square lattice
$\mathbb{Z}^2$} is a set of points of the lattice $\mathbb{Z}^2$
with coordinates $\{(i + s, j + r), 0 \leq s, r \leq N - 1\}$,
where $i, j$ is an arbitrary f\/ixed pair of integers, $i, j \in
\mathbb{Z}$, $N \geq 2$.

Let us consider a discrete equation on $\mathbb{Z}^2$ def\/ined by a
relation for the values of the f\/ield $u$ at the points of the
lattice $\mathbb{Z}^2$ that form {\it elementary $3 \times 3$
squares}:
\begin{gather} Q (u_{i j}, \ldots, u_{i + s, j + r}, \ldots, u_{i +
2, j + 2}) = 0, \qquad  0 \leq s, r \leq 2, \quad i, j \in \mathbb{Z},
\label{4}\end{gather}
so that in each elementary $3 \times 3$ square of the
lattice $\mathbb{Z}^2$, i.e., in each set of lattice points with
coordinates of the form $\{(i, j), (i + 1, j), (i + 2, j), (i, j +
1), (i + 1, j + 1), (i + 2, j + 1), (i, j + 2)$, $(i + 1, j + 2), (i +
2, j + 2)\}$, $i, j \in \mathbb{Z}$, the value of the f\/ield $u$ at
one of the points of this elementary $3 \times 3$ square is
determined by the values of the f\/ield at the other eight points.

For def\/initeness, one can require, for example, that relations~\eqref{4}
are invariant with respect to the full symmetry group of
the conf\/iguration of points of the lattice $\mathbb{Z}^2$ that form
elementary $3 \times 3$ squares (obviously, this group of symmetries
coincides with the full symmetry group of the usual square). For any
discrete equation of the form~\eqref{4}, f\/ixing generic initial
data, for example, on two bands along the coordinate axes of the
lattice $\mathbb{Z}^2$, $\{(i, 0), (i, 1), i \in \mathbb{Z} \}$ and
$\{(0, j), (1, j), j \in \mathbb{Z} \}$, completely determines a
f\/ield $u$ on $\mathbb{Z}^2$ that satisf\/ies this equation.

Quite similarly, discrete equations on $\mathbb{Z}^2$ given by
relations for the values of the f\/ield $u$ at the points of the
lattice $\mathbb{Z}^2$ that form elementary $N \times N$ squares can
be def\/ined for an arbitrary $N \geq 2$. For def\/initeness, one can
again require, for example, that the relations are invariant with
respect to the full symmetry group of the conf\/iguration of points of
the lattice $\mathbb{Z}^2$ that form elementary $N \times N$ squares
(moreover, it is also obvious that for any $N \geq 2$ the full
symmetry group of the conf\/iguration of points of the lattice
$\mathbb{Z}^2$ that form elementary $N \times N$ squares coincides
with the full symmetry group of the usual square).

We consider the cubic lattice $\mathbb{Z}^3$ and {\it the
consistency condition for discrete equations of the form~\eqref{4}
on all two-dimensional coordinate sublattices of the
cubic lattice $\mathbb{Z}^3$}. Initial data can be specif\/ied, for
example, at the following lattice points that are situated on 12
straight lines going along the coordinate axes of the lattice
$\mathbb{Z}^3$: $(i, 0, 0)$, $(i, 1, 0)$, $(i, 0, 1)$, $(i, 1, 1)$,
$(0, j, 0)$, $(1, j, 0)$, $(0, j, 1)$, $(1, j, 1)$, $(0, 0, k)$,
$(1, 0, k)$, $(0, 1, k)$, and $(1, 1, k)$, $i, j, k \in \mathbb{Z}$.
The values of the f\/ield $u$ at all other points of the cubic lattice
$\mathbb{Z}^3$ must then be correctly determined in a consistent way
by relations~\eqref{4} on all elementary $3 \times 3$ squares of all
two-dimensional coordinate sublattices of the cubic lattice
$\mathbb{Z}^3$ for generic initial data. In the elementary cube
$\{(i, j, k), 0 \leq i, j, k \leq 2 \}$ the values $u_{2 0 2}$,
$u_{2 1 2}$, $u_{2 2 0}$, $u_{2 2 1}$, $u_{0 2 2}$, and $u_{1 2 2}$
are determined using relations~\eqref{4} on the corresponding
elementary $3 \times 3$ squares situated in the cube under
consideration (three faces of the cube that are situated on the coordinate planes
and three middle normal sections of the
cube) by the initial data $u_{0 0 0}$, $u_{1 0 0}$, $u_{2 0 0}$,
$u_{0 1 0}$, $u_{1 1 0}$, $u_{2 1 0}$, $u_{0 0 1}$, $u_{1 0 1}$,
$u_{2 0 1}$, $u_{0 1 1}$, $u_{1 1 1}$, $u_{2 1 1}$, $u_{0 2 0}$,
$u_{1 2 0}$, $u_{0 2 1}$, $u_{1 2 1}$, $u_{0 0 2}$, $u_{1 0 2}$,
$u_{0 1 2}$, and $u_{1 1 2}$, and three relations must hold
simultaneously on three other faces of the cube for the value $u_{2 2 2}$.
One can assume that the consistency condition on the cubic lattice
$\mathbb{Z}^3$ for any discrete equation of the form~\eqref{4} is
the consistency condition of the corresponding overdetermined system
of relations on the value $u_{2 2 2}$ for generic initial data.

Quite similarly, for an arbitrary  $N \geq 2$, one can def\/ine {\it
the consistency condition on the cubic lattice $\mathbb{Z}^3$ for
discrete equations on $\mathbb{Z}^2$ given by relations on the
values of the field $u$ at the points of the lattice $\mathbb{Z}^2$
that form elementary $N \times N$ squares}. The values of the f\/ield
$u$ at all points of the cubic lattice $\mathbb{Z}^3$ must be
correctly determined in a consistent way by relations on all
elementary $N \times N$ squares of all two-dimensional coordinate
sublattices of the cubic lattice $\mathbb{Z}^3$ for generic initial
data. The consistency of discrete equations on the lattice
$\mathbb{Z}^2$ that are given by relations on the values of the
f\/ield $u$ at the points of the lattice $\mathbb{Z}^2$ that form
elementary $N \times N$ squares can be considered on one elementary
$N \times N \times N$ cube (the consistency of the relations on all
faces and all normal sections of the cube that are parallel to the
coordinate planes for generic initial data specif\/ied in the cube).

Let us consider the discrete nonlinear equation on
$\mathbb{Z}^2$ def\/ined by the condition that the determinants of all
$3 \times 3$ matrices of values of the f\/ield $u$ at the points of
the lattice $\mathbb{Z}^2$ that form elementary $3 \times 3$ squares
vanish:
\begin{gather}
u_{i, j + 2} u_{i + 1, j + 1} u_{i + 2, j} + u_{i, j
+ 1} u_{i + 1, j} u_{i + 2, j + 2} + u_{i, j} u_{i + 1, j +
2} u_{i + 2, j + 1}   \nonumber \\
{} - u_{i, j} u_{i + 1, j + 1} u_{i + 2, j + 2} - u_{i, j + 2} u_{i
+ 1, j} u_{i + 2, j + 1} - u_{i, j + 1} u_{i + 1, j + 2} u_{i + 2,
j} = 0, \qquad i, j \in \mathbb{Z}. \label{5}
\end{gather}
Equation~\eqref{5} is
linear with respect to each variable and invariant with respect to
the full symmetry group of the conf\/iguration of points of the
lattice $\mathbb{Z}^2$ that form elementary $3 \times 3$ squares.

It is not dif\/f\/icult to check that for generic initial data the
above-considered consistency condition on the cubic lattice is not
satisf\/ied for the discrete equation~\eqref{5}, and, in this sense,
the discrete nonlinear equation~\eqref{5} is not consistent on
two-dimensional coordinate sublattices of the cubic lattice
$\mathbb{Z}^3$.

We note that for such setting of the consistency problem on the
cubic lattice for discrete equations of the form~\eqref{4} we have
the following: in the elementary $3 \times 3 \times 3$ cube $\{(i,
j, k), 0 \leq i, j, k \leq 2 \}$ of the lattice $\mathbb{Z}^3$,
given initial data of values of the f\/ield $u$ at 20 points of this
elementary $3 \times 3 \times 3$ cube, one can determine the values
of the f\/ield $u$ at the other seven points of this elementary $3
\times 3 \times 3$ cube by relations~\eqref{4} (nine relations on
six faces and on three middle normal sections of the cube), and only
for the value of the f\/ield at one of the points one obtains an
overdetermined system consisting of three relations on three
distinct elementary $3 \times 3$ squares.

\section[Bent elementary $3 \times 3$ squares
and modified consistency conditions]{Bent elementary $\boldsymbol{3 \times 3}$ squares\\
and modif\/ied consistency conditions}

We consider a discrete equation of the form~\eqref{4} and require
that the discrete equation is sa\-tis\-f\/ied not only on all
two-dimensional coordinate sublattices of the cubic lattice
$\mathbb{Z}^3$, but also on all unions of two arbitrary intersecting
two-dimensional coordinate sublattices of the cubic lattice~$\mathbb{Z}^3$; i.e., the corresponding elementary $3 \times 3$
squares on which the discrete equation of the form~\eqref{4} is
considered can be {\it bent} at a right angle along any of two
middle lines of the elementary $3 \times 3$ square passing from one
two-dimensional coordinate sublattice to another so that all points
of {\it bent elementary $3 \times 3$ squares} are situated at
lattice points (one of lines of any bent elementary $3 \times 3$
square is situated on one of two intersecting two-dimensional
coordinate sublattices of the cubic lattice~$\mathbb{Z}^3$, one of
the lines is situated on the second of these two sublattices, and
the middle line which this elementary square is bent along is
situated on the intersection of these two intersecting
two-dimensional coordinate sublattices), for example, $\{(i, 0, 0),
(i, 1, 0), (i, 0, 1), i = 0, 1, 2\}$, $\{(0, j, 0), (1, j, 0), (0,
j, 1), j = 0, 1, 2\}$, and $\{(0, 0, k), (1, 0, k), (0, 1, k), k =
0, 1, 2\}$ ({\it bent elementary $3 \times 3$ squares}). We will
consider relation~\eqref{4} on all elementary $3 \times 3$ squares
(bent and unbent) all points of which are situated at lattice
points. In this case initial data can be specif\/ied, for example, at
the following points of the cubic lattice $\mathbb{Z}^3$: $(i, 0,
0)$, $(i, 0, 1)$, $(0, j, 0)$, $(0, j, 1)$, $(0, 0, k)$, $(1, 0,
k)$, $(1, 1, 0)$, and $(1, 1, 1)$, $i, j, k \in \mathbb{Z}$. The
values of the f\/ield $u$ at all other points of the cubic lattice
$\mathbb{Z}^3$ must then be correctly determined in a consistent way
by relations~\eqref{4} on all elementary $3 \times 3$ squares
(including all bent elementary $3 \times 3$ squares) of all unions
of two arbitrary intersecting two-dimensional coordinate sublattices
of the cubic lattice~$\mathbb{Z}^3$ for generic initial data. In the
elementary cube $\{(i, j, k), 0 \leq i, j, k \leq 2 \}$ the values
$u_{0 1 2}$, $u_{0 2 2}$, $u_{1 1 2}$, $u_{2 0 2}$, $u_{1 2 k}$,
$u_{2 1 k}$, and $u_{2 2 k}$, $0 \leq k \leq 2$, are determined by
the initial data $u_{0 0 0}$, $u_{1 0 0}$, $u_{2 0 0}$, $u_{0 0 1}$,
$u_{1 1 0}$, $u_{0 1 0}$, $u_{1 0 1}$, $u_{2 0 1}$, $u_{1 1 1}$,
$u_{0 1 1}$, $u_{0 0 2}$, $u_{0 2 0}$, $u_{0 2 1}$, and $u_{1 0 2}$
and by overdetermined systems generated by relations~\eqref{4} on
elementary $3 \times 3$ squares (including all bent elementary $3
\times 3$ squares) that are situated in the cube under consideration
(six faces, three middle normal sections of the cube, and 48 bent
elementary $3 \times 3$ squares). There are 48 distinct bent
elementary $3 \times 3$ squares in any elementary $3 \times 3 \times
3$ cube, which can be easily counted by the bending edges of the
bent elementary $3 \times 3$ squares: to each of the 12 edges of the
cube, there corresponds one bent elementary $3 \times 3$ square in
the cube; to each of the 12 middle lines of points on the faces of
the cube ($2 \times 6$), there correspond two distinct bent
elementary $3 \times 3$ squares in the cube; and to each of the
three interior lines of points in the cube that connect the centres
of opposite faces of the cube, there correspond four distinct bent
elementary $3 \times 3$ squares in the cube. One can assume that the
consistency condition on the cubic lattice~$\mathbb{Z}^3$ for any
discrete equation of the form~\eqref{4} (the global consistency) is the consistency condition
of the corresponding overdetermined system of relations on the
values of the f\/ield $u$ in the elementary cube $\{(i, j, k), 0
\leq i, j, k \leq 2 \}$ for generic initial data (the local consistency). The corresponding
discrete equations will also be called {\it consistent on the cubic
lattice}.

We note that for this new setting of the consistency problem on the
cubic lattice for discrete equations of the form~\eqref{4} we have
the following: in the elementary $3 \times 3 \times 3$ cube $\{(i,
j, k),  0 \leq i, j, k \leq 2 \}$ of the lattice $\mathbb{Z}^3$,
given initial data of values of the f\/ield $u$ at 14 points of this
elementary $3 \times 3 \times 3$ cube, one can determine the values
of the f\/ield $u$ at the other 13 points of this elementary $3 \times
3 \times 3$ cube by relations~\eqref{4} (57 relations on six faces,
on three middle normal sections of the cube, and on 48 bent
elementary $3 \times 3$ squares), which constitute in this case a~highly overdetermined system of relations.

We also note that the global consistency follows
from the local consistency, i.e., if the consistency condition is fulf\/illed
for generic initial data in the elementary cube $\{(i, j, k), 0
\leq i, j, k \leq 2 \}$, then it is fulf\/illed for generic initial data everywhere in
the cubic lattice $\mathbb{Z}^3$, i.e., in each elementary $3 \times
3 \times 3$ cube $\{(i_0 + i, j_0 + j, k_0 + k), 0
\leq i, j, k \leq 2 \}$, $i_0, j_0, k_0 \in \mathbb{Z}$.
We give here a formal scheme of proof. The number of given initial data in an arbitrary elementary $3 \times
3 \times 3$ cube $\{(i_0 + i, j_0 + j, k_0 + k), 0
\leq i, j, k \leq 2 \}$, $i_0, j_0, k_0 \in \mathbb{Z}$, is not more than 14 as we have in the ``main''
elementary cube $\{(i, j, k), 0
\leq i, j, k \leq 2 \}$. We start from the ``main''
elementary cube $\{(i, j, k), 0
\leq i, j, k \leq 2 \}$, where the consistency condition is fulf\/illed for generic initial data as the local consistency by our assumption, and then we will consider some special shifts of elementary $3 \times
3 \times 3$ cubes step by step from the ``main''
elementary cube $\{(i, j, k), 0
\leq i, j, k \leq 2 \}$ until f\/ill all the cubic lattice $\mathbb{Z}^3$. There are three essentially dif\/ferent situations, when after shifting we obtain new 9 points of the cubic lattice $\mathbb{Z}^3$ (the face of the shifted elementary $3 \times
3 \times 3$ cube), new 3 points (the edge of the shifted elementary $3 \times
3 \times 3$ cube) or new only one point (the vertice of the shifted elementary $3 \times
3 \times 3$ cube). First of all, we consider 6 shifted elementary $3 \times
3 \times 3$ cubes (6 elementary $3 \times
3 \times 3$ cubes shifted to the side of each of the faces of the ``main''
elementary cube $\{(i, j, k), 0
\leq i, j, k \leq 2 \}$). In these elementary $3 \times
3 \times 3$ cubes we have some given initial data (not more than 14) and some values of the f\/ield $u$ determined in the
``main'' elementary cube $\{(i, j, k), 0
\leq i, j, k \leq 2 \}$. Let us consider any of these elementary $3 \times
3 \times 3$ cubes and prove that the consistency condition is also fulf\/illed in it. We consider some determined values as new initial data for this elementary $3 \times
3 \times 3$ cube, f\/irst of all, we take given initial data and if it is necessary (i.e., if the number of given initial data in this elementary $3 \times
3 \times 3$ cube is less than 14), add to given initial data the corresponding values determined in the
``main'' elementary cube $\{(i, j, k), 0
\leq i, j, k \leq 2 \}$. It is simple to check that we always can do this. Then we can determine all values in this elementary $3 \times
3 \times 3$ cube in a consistent way by our assumption of the local consistency. After that we must prove that the values determined simultaneously from the ``main'' elementary cube $\{(i, j, k), 0
\leq i, j, k \leq 2 \}$ and from the elementary $3 \times
3 \times 3$ cube under consideration coincide. It follows from the fact that all these values can be determined step by step from elementary $3 \times 3$ squares (bent and unbent) situated in both these elementary $3 \times
3 \times 3$ cubes simultaneously. Similarly, we prove step by step that the consistency condition is fulf\/illed in
any elementary $3 \times
3 \times 3$ cube obtained from the ``main'' elementary cube $\{(i, j, k), 0
\leq i, j, k \leq 2 \}$ by a shift along the coordinate axes (along the given bands of initial data).
Then we prove step by step that the consistency condition is fulf\/illed in
any elementary $3 \times
3 \times 3$ cube obtained from the considered elementary $3 \times
3 \times 3$ cubes by shifts in the corresponding coordinate planes. This is the second type of our shifts. Again we must consider some initial data in each elementary $3 \times
3 \times 3$ cube and determine all values in it in a consistent way by our assumption of the local consistency. All
the values determined simultaneously from dif\/ferent elementary $3 \times
3 \times 3$ cubes coincide since all these values can be determined step by step from elementary $3 \times 3$ squares (bent and unbent) situated in both these elementary $3 \times
3 \times 3$ cubes simultaneously. Similarly, we prove step by step that the consistency condition is fulf\/illed for the shifts of the third type f\/illing all the cubic lattice $\mathbb{Z}^3$.

Quite similarly, for an arbitrary $N > 2$, one can def\/ine the
corresponding {\it modified consistency condition on the cubic lattice
$\mathbb{Z}^3$ for discrete equations on the square lattice
$\mathbb{Z}^2$ that are given by relations on the values of the field
$u$ at the points of the lattice $\mathbb{Z}^2$ that form elementary
$N \times N$ squares {\rm (}including any bent elementary $N \times
N$ squares{\rm )}}. Moreover, for $N > 3$, one can, generally
speaking, allow a larger number (up to $N - 2$) of bendings
of elementary $N \times N$ squares in the cubic lattice
$\mathbb{Z}^3$ (in this case each elementary $N \times N$ square can
be bent in the cubic lattice simultaneously along up to $N -
2$ parallel lines of the same type, each bending being to one of the
two possible dif\/ferent sides).

\section[Consistency on cubic lattices for $3 \times 3$ determinants]{Consistency on cubic lattices for $\boldsymbol{3 \times 3}$ determinants}

The following basic theorem holds.

\begin{theorem}[\cite{1}] \label{theorem1}
For arbitrary generic initial data, the
nonlinear discrete equation~\eqref{5} can be satisfied in a
consistent way on all unions of pairs of arbitrary intersecting
two-dimensional coordinate sublattices of the cubic lattice
$\mathbb{Z}^3$; i.e., the discrete nonlinear equation~\eqref{5}
is consistent on the cubic lattice $\mathbb{Z}^3$.
\end{theorem}

\begin{proof}\looseness=-1
 Here we give the strict proof of consistency
``around the elementary cube'' (the local consistency) for generic initial data.
Let us consider the elementary cube $\{(i, j, k), 0
\leq i, j, k \leq 2 \}$ of the cubic lattice $\mathbb{Z}^3$ and
specify generic initial data at the following points of this
elementary cube: $(i, 0, 0)$, $(i, 0, 1)$, $(0, j, 0)$, $(0, j, 1)$,
$(0, 0, k)$, $(1, 0, k)$, $(1, 1, 0)$, and $(1, 1, 1)$, $0 \leq i,
j, k \leq 2$. In particular, it is suf\/f\/icient to require that the
following condition on initial data is fulf\/illed: in all elementary
$3 \times 3$ squares (bent and unbent) situated in the elementary
cube under consideration, all $2 \times 2$ minors that are
completely formed by only initial data are not equal to zero (this
is a~condition of general position for initial data). In this paper,
we will assume that precisely this concrete condition of general
position on initial data is fulf\/illed. We note that the condition on
initial data, when no four values $u_a$, $u_b$, $u_c$, $u_d$ of initial
data from distinct points $a$, $b$, $c$, $d$ of the cube under
consideration satisfy the relation $u_a u_b = u_c u_d$ (i.e., if we
arrange values of initial data from distinct points of the cube
under consideration in all points of elementary $2 \times 2$ square
in an arbitrary way, then all the determinants of the corresponding
$2 \times 2$ matrices obtained by this procedure are not equal to
zero), is also a condition of general position for initial data. We
require the fulf\/illment of a weaker condition on initial data, when
only some of these determinants, namely, only those from them that
are $2 \times 2$ minors formed by initial data in the $3 \times 3$
matrices of values of the f\/ield at the points of the elementary $3
\times 3$ squares (bent and unbent) situated in the elementary cube
under consideration, must not be equal to zero. We will distinguish
the following three dif\/ferent types of lines of lattice points in
the elementary cube under consideration: the lines parallel to the
$x$-axis, i.e., the sets of points of the form $\{(i, r, s), 0 \leq
i \leq 2\}$, where $(r, s)$ are f\/ixed ordered pairs of integers, $0
\leq r, s \leq 2$, that number the lines in this elementary cube
that are parallel to the $x$-axis ({\it the $x$-type lines}); the
lines parallel to the $y$-axis, i.e., the sets of points of the form
$\{(r, j, s), 0 \leq j \leq 2\}$, where $(r, s)$ are f\/ixed ordered
pairs of integers, $0 \leq r, s \leq 2$, that number the lines in
the elementary cube under consideration that are parallel to the
$y$-axis ({\it the $y$-type lines}); and the lines parallel to the
$z$-axis, i.e., the sets of points of the form $\{(r, s, k), 0 \leq
k \leq 2\}$, where $(r, s)$ are f\/ixed ordered pairs of integers, $0
\leq r, s \leq 2$, that number the lines in the elementary cube
under consideration that are parallel to the $z$-axis ({\it the
$z$-type lines}). The specif\/ied initial data f\/ill a pair of lines of
each of these three types (a pair of lines parallel to the
corresponding coordinate axis for each of the coordinate axes). We
will consider all these lines as {\it basic} ones: $\{(i, 0, 0), 0
\leq i \leq 2\}$ and $\{(i, 0, 1), 0 \leq i \leq 2\}$ ({\it the
basic $x$-type lines}); $\{(0, j, 0), 0 \leq j \leq 2\}$ and $\{(0,
j, 1), 0 \leq j \leq 2\}$ ({\it the basic $y$-type lines}); and
$\{(0, 0, k), 0 \leq k \leq 2\}$ and $\{(1, 0, k), 0 \leq k \leq
2\}$ ({\it the basic $z$-type lines}). The vectors of values of the
f\/ield $u$ at the points of the basic lines will be called {\it basic
vectors} (of the corresponding type). Note that the vectors of
values of the f\/ield $u$ at the points of the basic lines of each
type are linearly independent, since otherwise the corresponding $2
\times 2$ minors formed by initial data must be equal to zero.
Therefore, basic vectors of each type are linearly independent.
Given arbitrary generic initial data, we will determine the values
of the scalar f\/ield $u$ at the remaining points of the elementary
cube under consideration according to relations~\eqref{5} and mark
the points at which the values of the f\/ield have already been found.
We will also shade each line in this elementary cube if the vector
of values of the f\/ield~$u$ at the points of this line is a linear
combination of the vectors of values of the f\/ield~$u$ at the points
of the two basic lines of the same type (for the coordinates of
vectors of values of the f\/ield~$u$ at the points of lines of the
same type, there is a natural ascending order of the respective
coordinate~$x$,~$y$ or~$z$). First of all, in the elementary cube
under consideration we must mark all lattice points at which the
initial data are given and shade all basic lines of all three types
by the very def\/inition of this procedure. It is obvious that if
carrying out such a procedure for generic initial data yields all
the lattice points marked and all the lines of all the types shaded
in the elementary cube under consideration, then the theorem will be
proved, because in this case, {\it for any three lines of the same
type in this elementary cube} (and, hence, for any elementary $3
\times 3$ square in this elementary cube, bent or unbent), the
determinant of the matrix of values of the scalar f\/ield $u$ at the
points of these lines will vanish, and this is even more than is
required for the consistency of the corresponding discrete equation.
Thus, in this case, as a matter of fact, we will prove even a {\it
considerably stronger principle of consistency on the cubic lattice
$\mathbb{Z}^3$ for determinants and for the nonlinear discrete
equation~\eqref{5}}. It remains to shade all the lines of all
the types in the elementary cube under consideration. For this
purpose, it is necessary to consider consecutively at least 13
elementary $3 \times 3$ squares (bent and unbent) of our elementary
cube determining values of the f\/ield $u$ at 13 unmarked points.

Let us consider the elementary $3 \times 3$ square $\{(i, 0, 0), (i,
0, 1), (i, 0, 2), i = 0, 1, 2\}$ in our cube (a~face of the cube).
In this elementary square the values of the f\/ield $u$ are given at
eight points and the value of the f\/ield $u$ at the remaining ninth
point $(2, 0, 2)$ is determined by relation~\eqref{5}, i.e.,
by the condition that the determinant of the matrix of values of the
f\/ield at the lattice points of this elementary $3 \times 3$ square
vanishes, because the corresponding $2 \times 2$ minor formed by
initial data is not equal to zero. Since basic vectors of the same
type are linearly independent, the vector of values of the f\/ield $u$
at the points of the line $\{(i, 0, 2), 0 \leq i \leq 2\}$ is a
linear combination of the vectors of values of the f\/ield $u$ at the
points of the two basic lines of the same type, $\{(i, 0, 0), 0 \leq
i \leq 2\}$ and $\{(i, 0, 1), 0 \leq i \leq 2\}$, situated in the
given elementary $3 \times 3$ square; i.e., we can mark the point
$(2, 0, 2)$ and shade the line $\{(i, 0, 2), 0 \leq i \leq 2\}$.
Moreover, the obtained vector of values of the f\/ield $u$ at the
points of the line $\{(i, 0, 2), 0 \leq i \leq 2\}$ forms a linearly
independent pair of vectors with each of the basic vectors of the
same type, since otherwise the corresponding $2 \times 2$ minors
formed by initial data must be equal to zero.

Similarly, since basic vectors of the same type are linearly
independent, it follows imme\-diately from vanishing the determinant
of the matrix of values of the f\/ield at the lattice points of this
elementary $3 \times 3$ square that the vector of values of the
f\/ield $u$ at the points of the line $\{(2, 0, k), 0 \leq k \leq 2\}$
is a linear combination of the vectors of values of the f\/ield $u$ at
the points of the other two lines of this elementary $3 \times 3$
square, namely, the two basic lines of the same type, $\{(0, 0, k),
0 \leq k \leq 2\}$ and $\{(1, 0, k), 0 \leq k \leq 2\}$, situated in
the given elementary $3 \times 3$ square; i.e., we can shade the
line $\{(2, 0, k), 0 \leq k \leq 2\}$. Moreover, the obtained vector
of values of the f\/ield $u$ at the points of the line $\{(2, 0, k), 0
\leq k \leq 2\}$ forms a linearly independent pair of vectors with
each of the basic vectors of the same type, since otherwise the
corresponding $2 \times 2$ minors formed by initial data must be
equal to zero.

Let us consider the bent elementary $3 \times 3$ square $\{(1, 0,
k), (0, 0, k), (0, 1, k), k = 0, 1, 2\}$ in our cube. In this
elementary square the values of the f\/ield $u$ are given at eight
points and the value of the f\/ield $u$ at the remaining ninth point
$(0, 1, 2)$ is determined by relation~\eqref{5}, i.e., by the
condition that the determinant of the matrix of values of the f\/ield
at the lattice points of this bent elementary $3 \times 3$ square
vanishes, because the corresponding $2 \times 2$ minor formed by
initial data is not equal to zero. Since basic vectors of the same
type are linearly independent, the vector of values of the f\/ield $u$
at the points of the line $\{(0, 1, k), 0 \leq k \leq 2\}$ is a
linear combination of the vectors of values of the f\/ield $u$ at the
points of the two basic lines of the same type, $\{(0, 0, k), 0 \leq
k \leq 2\}$ and $\{(1, 0, k), 0 \leq k \leq 2\}$, situated in the
given bent elementary $3 \times 3$ square; i.e., we can mark the
point $(0, 1, 2)$ and shade the line $\{(0, 1, k), 0 \leq k \leq
2\}$. Moreover, the obtained vector of values of the f\/ield $u$ at
the points of the line $\{(0, 1, k), 0 \leq k \leq 2\}$ forms a
linearly independent pair of vectors with each of the basic vectors
of the same type, since otherwise the corresponding $2 \times 2$
minors formed by initial data must be equal to zero.

Now we consider another bent elementary $3 \times 3$ square $\{(1,
1, k), (1, 0, k), (0, 0, k), k = 0, 1, 2\}$ in our cube. In this
elementary square the values of the f\/ield $u$ are given at eight
points and the value of the f\/ield $u$ at the remaining ninth point
$(1, 1, 2)$ is determined by relation~\eqref{5}, i.e., by the
condition that the determinant of the matrix of values of the f\/ield
at the lattice points of this bent elementary $3 \times 3$ square
vanishes, because the corresponding $2 \times 2$ minor formed by
initial data is not equal to zero. Since basic vectors of the same
type are linearly independent, the vector of values of the f\/ield $u$
at the points of the line $\{(1, 1, k), 0 \leq k \leq 2\}$ is a
linear combination of the vectors of values of the f\/ield $u$ at the
points of the two basic lines of the same type, $\{(1, 0, k), 0 \leq
k \leq 2\}$ and $\{(0, 0, k), 0 \leq k \leq 2\}$, situated in the
given bent elementary $3 \times 3$ square; i.e., we can mark the
point $(1, 1, 2)$ and shade the line $\{(1, 1, k), 0 \leq k \leq
2\}$. Moreover, the obtained vector of values of the f\/ield $u$ at
the points of the line $\{(1, 1, k), 0 \leq k \leq 2\}$ forms a
linearly independent pair of vectors with each of the basic vectors
of the same type, since otherwise the corresponding $2 \times 2$
minors formed by initial data must be equal to zero.

Let us consider one more bent elementary $3 \times 3$ square $\{(i,
0, 0), (i, 0, 1), (i, 1, 1), i = 0, 1, 2\}$ in our cube. In this
elementary square the values of the f\/ield $u$ are given at eight
points and the value of the f\/ield $u$ at the remaining ninth point
$(2, 1, 1)$ is determined by relation~\eqref{5}, i.e., by the
condition that the determinant of the matrix of values of the f\/ield
at the lattice points of this bent elementary $3 \times 3$ square
vanishes, because the corresponding $2 \times 2$ minor formed by
initial data is not equal to zero. Since basic vectors of the same
type are linearly independent, the vector of values of the f\/ield $u$
at the points of the line $\{(i, 1, 1), 0 \leq i \leq 2\}$ is a
linear combination of the vectors of values of the f\/ield $u$ at the
points of the two basic lines of the same type, $\{(i, 0, 0), 0 \leq
i \leq 2\}$ and $\{(i, 0, 1), 0 \leq i \leq 2\}$, situated in the
given bent elementary $3 \times 3$ square; i.e., we can mark the
point $(2, 1, 1)$ and shade the line $\{(i, 1, 1), 0 \leq i \leq
2\}$. Moreover, the obtained vector of values of the f\/ield $u$ at
the points of the line $\{(i, 1, 1), 0 \leq i \leq 2\}$ forms a
linearly independent pair of vectors with each of the basic vectors
of the same type, since otherwise the corresponding $2 \times 2$
minors formed by initial data must be equal to zero.

Let us consider the next bent elementary $3 \times 3$ square $\{(i,
0, 1), (i, 0, 0), (i, 1, 0),  i = 0, 1, 2\}$ in our cube. In this
elementary square the values of the f\/ield $u$ are given at eight
points and the value of the f\/ield $u$ at the remaining ninth point
$(2, 1, 0)$ is determined by relation~\eqref{5}, i.e., by the
condition that the determinant of the matrix of values of the f\/ield
at the lattice points of this bent elementary $3 \times 3$ square
vanishes, because the corresponding $2 \times 2$ minor formed by
initial data is not equal to zero. Since basic vectors of the same
type are linearly independent, the vector of values of the f\/ield $u$
at the points of the line $\{(i, 1, 0), 0 \leq i \leq 2\}$ is a
linear combination of the vectors of values of the f\/ield $u$ at the
points of the two basic lines of the same type, $\{(i, 0, 0), 0 \leq
i \leq 2\}$ and $\{(i, 0, 1), 0 \leq i \leq 2\}$, situated in the
given bent elementary $3 \times 3$ square; i.e., we can mark the
point $(2, 1, 0)$ and shade the line $\{(i, 1, 0), 0 \leq i \leq
2\}$. Moreover, the obtained vector of values of the f\/ield $u$ at
the points of the line $\{(i, 1, 0), 0 \leq i \leq 2\}$ forms a
linearly independent pair of vectors with each of the basic vectors
of the same type, since otherwise the corresponding $2 \times 2$
minors formed by initial data must be equal to zero.

Let us consider one more bent elementary $3 \times 3$ square $\{(1,
j, 0), (0, j, 0), (0, j, 1),  j = 0, 1, 2\}$ in our cube. In this
elementary square the values of the f\/ield $u$ are given at eight
points and the value of the f\/ield $u$ at the remaining ninth point
$(1, 2, 0)$ is determined by relation~\eqref{5}, i.e., by the
condition that the determinant of the matrix of values of the f\/ield
at the lattice points of this bent elementary $3 \times 3$ square
vanishes, because the corresponding $2 \times 2$ minor formed by
initial data is not equal to zero. Since basic vectors of the same
type are linearly independent, the vector of values of the f\/ield $u$
at the points of the line $\{(1, j, 0), 0 \leq j \leq 2\}$ is a
linear combination of the vectors of values of the f\/ield $u$ at the
points of the two basic lines of the same type, $\{(0, j, 0), 0 \leq
j \leq 2\}$ and $\{(0, j, 1), 0 \leq j \leq 2\}$, situated in the
given bent elementary $3 \times 3$ square; i.e., we can mark the
point $(1, 2, 0)$ and shade the line $\{(1, j, 0), 0 \leq j \leq
2\}$. Moreover, the obtained vector of values of the f\/ield $u$ at
the points of the line $\{(1, j, 0), 0 \leq j \leq 2\}$ forms a
linearly independent pair of vectors with each of the basic vectors
of the same type, since otherwise the corresponding $2 \times 2$
minors formed by initial data must be equal to zero.

Let us consider the next bent elementary $3 \times 3$ square $\{(0,
j, 0), (0, j, 1), (1, j, 1), j = 0, 1, 2\}$ in our cube. In this
elementary square the values of the f\/ield $u$ are given at eight
points and the value of the f\/ield $u$ at the remaining ninth point
$(1, 2, 1)$ is determined by relation~\eqref{5}, i.e., by the
condition that the determinant of the matrix of values of the f\/ield
at the lattice points of this bent elementary $3 \times 3$ square
vanishes, because the corresponding $2 \times 2$ minor formed by
initial data is not equal to zero. Since basic vectors of the same
type are linearly independent, the vector of values of the f\/ield $u$
at the points of the line $\{(1, j, 1), 0 \leq j \leq 2\}$ is a
linear combination of the vectors of values of the f\/ield $u$ at the
points of the two basic lines of the same type, $\{(0, j, 0), 0 \leq
j \leq 2\}$ and $\{(0, j, 1), 0 \leq j \leq 2\}$, situated in the
given bent elementary $3 \times 3$ square; i.e., we can mark the
point $(1, 2, 1)$ and shade the line $\{(1, j, 1), 0 \leq j \leq
2\}$. Moreover, the obtained vector of values of the f\/ield $u$ at
the points of the line $\{(1, j, 1), 0 \leq j \leq 2\}$ forms a
linearly independent pair of vectors with each of the basic vectors
of the same type, since otherwise the corresponding $2 \times 2$
minors formed by initial data must be equal to zero.

Note that the vectors of values of the f\/ield $u$ at the points of
the shaded lines $\{(1, j, 0), 0 \leq j \leq 2\}$ and $\{(1, j, 1),
0 \leq j \leq 2\}$, $\{(i, 1, 0), 0 \leq i \leq 2\}$ and $\{(i, 1,
1), 0 \leq i \leq 2\}$, $\{(0, 1, k), 0 \leq k \leq 2\}$ and $\{(1,
1, k), 0 \leq k \leq 2\}$, $\{(2, 0, k), 0 \leq k \leq 2\}$ and
$\{(1, 1, k), 0 \leq k \leq 2\}$, are linearly independent, since
otherwise the corresponding $2 \times 2$ minors formed by initial
data must be equal to zero.

Let us consider one more elementary $3 \times 3$ square $\{(0, j,
0), (0, j, 1), (0, j, 2), j = 0, 1, 2\}$ in our cube (a face of the
cube). In this elementary square at the present moment the values of
the f\/ield $u$ are already determined at eight points and the value
of the f\/ield $u$ at the remaining ninth point $(0, 2, 2)$ is
determined by relation~\eqref{5}, i.e., by the condition that
the determinant of the matrix of values of the f\/ield at the lattice
points of this elementary $3 \times 3$ square vanishes, because the
corresponding $2 \times 2$ minor formed by initial data is not equal
to zero. Since basic vectors of the same type are linearly
independent, the vector of values of the f\/ield $u$ at the points of
the line $\{(0, j, 2), 0 \leq j \leq 2\}$ is a linear combination of
the vectors of values of the f\/ield $u$ at the points of the two
basic lines of the same type, $\{(0, j, 0), 0 \leq j \leq 2\}$ and
$\{(0, j, 1), 0 \leq j \leq 2\}$, situated in the given elementary
$3 \times 3$ square; i.e., we can mark the point $(0, 2, 2)$ and
shade the line $\{(0, j, 2), 0 \leq j \leq 2\}$. But in this case we
do not state that the obtained vector of values of the f\/ield $u$ at
the points of the line $\{(0, j, 2), 0 \leq j \leq 2\}$ forms
linearly independent pairs of vectors with basic vectors of the same
type.

\looseness=1
We note that if the vector of values of the f\/ield $u$ at the lattice
points of an arbitrary line is a~linear combination of the vectors
of values of the f\/ield $u$ at the lattice points of two shaded lines
of the same type, then this vector is a linear combination of the
vectors of values of the f\/ield $u$ at the points of the two basic
lines of the same type. This follows immediately from the fact that
each vector of values of the f\/ield $u$ at the points of an arbitrary
shaded line is a linear combination of the vectors of values of the
f\/ield $u$ at the points of the two basic lines of the same type.

\looseness=-1
Since the determinant of the matrix of values of the f\/ield at the
points of the elementary $3 \times 3$ square $\{(0, j, 0), (0, j,
1), (0, j, 2), j = 0, 1, 2\}$ (on a face of our cube) vanishes and,
as it was noted above, the vectors of values of the f\/ield $u$ at the
points of the two shaded lines, $\{(0, 0, k), 0 \leq k \leq 2\}$ and
$\{(0, 1, k), 0 \leq k \leq 2\}$, are linearly independent, it
follows immediately that the vector of values of the f\/ield $u$ at
the points of the line $\{(0, 2, k), 0 \leq k \leq 2\}$ is a~linear
combination of the vectors of values of the f\/ield $u$ at the points
of these two lines of this elementary $3 \times 3$ square, namely,
two shaded lines of the same type, $\{(0, 0, k), 0 \leq k \leq 2\}$
and $\{(0, 1, k), 0 \leq k \leq 2\}$, situated in the given
elementary $3 \times 3$ square; i.e., we can shade the line $\{(0,
2, k), 0 \leq k \leq 2\}$. Moreover, the obtained vector of values
of the f\/ield $u$ at the points of the line $\{(0, 2, k), 0 \leq k
\leq 2\}$ forms a~linearly independent pair of vectors with the
vector of values of the f\/ield $u$ at the points of the basic line
$\{(0, 0, k), 0 \leq k \leq 2\}$ and also with the vector of values
of the f\/ield $u$ at the points of the line $\{(0, 1, k), 0 \leq k
\leq 2\}$ and the vector of values of the f\/ield $u$ at the points of
the line $\{(1, 1, k), 0 \leq k \leq 2\}$, since otherwise the
corresponding $2 \times 2$ minors formed by initial data must be
equal to zero.

In the elementary $3 \times 3$ square $\{(i, 0, 0), (i, 1, 0), (i,
2, 0), i = 0, 1, 2\}$ of our cube (on a face of the cube) at the
present moment the values of the f\/ield $u$ are already determined at
eight points and the value of the f\/ield $u$ at the remaining ninth
point $(2, 2, 0)$ is determined by relation~\eqref{5}, i.e.,
by the condition that the determinant of the matrix of values of the
f\/ield at the points of this elementary $3 \times 3$ square vanishes,
because the corresponding $2 \times 2$ minor formed by initial data
is not equal to zero. Since, as it was noted above, the vectors of
values of the f\/ield $u$ at the points of the two shaded lines,
$\{(i, 0, 0), 0 \leq i \leq 2\}$ and $\{(i, 1, 0), 0 \leq i \leq
2\}$, are linearly independent, the vector of values of the f\/ield
$u$ at the points of the line $\{(i, 2, 0), 0 \leq i \leq 2\}$ is a~linear combination of the vectors of values of the f\/ield $u$ at the
points of these two shaded lines of the same type, $\{(i, 0, 0), 0
\leq i \leq 2\}$ and $\{(i, 1, 0), 0 \leq i \leq 2\}$, situated in
the given elementary $3 \times 3$ square; i.e., we can mark the
point $(2, 2, 0)$ and shade the line $\{(i, 2, 0), 0 \leq i \leq
2\}$. But in this case we do not state that the obtained vector of
values of the f\/ield $u$ at the points of the line $\{(i, 2, 0), 0
\leq i \leq 2\}$ forms linearly independent pairs of vectors with
other vectors of values of the f\/ield $u$ at the points of shaded
lines of the same type.

Since the determinant of the matrix of values of the f\/ield at the
points of the elementary $3 \times 3$ square $\{(i, 0, 0), (i, 1,
0), (i, 2, 0), i = 0, 1, 2\}$ (on a face of our cube) vanishes and,
as it was noted above, the vectors of values of the f\/ield $u$ at the
points of the two shaded lines, $\{(0, j, 0), 0 \leq j \leq 2\}$ and
$\{(1, j, 0), 0 \leq j \leq 2\}$, are linearly independent, it
follows immediately that the vector of values of the f\/ield $u$ at
the points of the line $\{(2, j, 0), 0 \leq j \leq 2\}$ is a~linear
combination of the vectors of values of the f\/ield $u$ at the points
of two other lines of this elementary $3 \times 3$ square, namely,
two shaded lines of the same type, $\{(0, j, 0), 0 \leq j \leq 2\}$
and $\{(1, j, 0), 0 \leq j \leq 2\}$, situated in the given
elementary $3 \times 3$ square; i.e., we can shade the line $\{(2,
j, 0), 0 \leq j \leq 2\}$. But in this case we do not state that the
obtained vector of values of the f\/ield $u$ at the points of the line
$\{(2, j, 0), 0 \leq j \leq 2\}$ forms linearly independent pairs of
vectors with other vectors of values of the f\/ield $u$ at the points
of shaded lines of the same type.

In the elementary $3 \times 3$ square $\{(i, 0, 1), (i, 1, 1), (i,
2, 1), i = 0, 1, 2\}$ of our cube (on a middle normal section of the
cube) at the present moment the values of the f\/ield $u$ are already
determined at eight points and the value of the f\/ield $u$ at the
remaining ninth point $(2, 2, 1)$ is determined by relation~\eqref{5},
i.e., by the condition that the determinant of the
matrix of values of the f\/ield at the points of this elementary $3
\times 3$ square vanishes, because the corresponding $2 \times 2$
minor formed by initial data is not equal to zero. Since, as it was
noted above, the vectors of values of the f\/ield $u$ at the points of
the two shaded lines, $\{(i, 0, 1), 0 \leq i \leq 2\}$ and $\{(i, 1,
1), 0 \leq i \leq 2\}$, are linearly independent, the vector of
values of the f\/ield $u$ at the points of the line $\{(i, 2, 1), 0
\leq i \leq 2\}$ is a~linear combination of the vectors of values of
the f\/ield $u$ at the points of these two shaded lines of the same
type, $\{(i, 0, 1), 0 \leq i \leq 2\}$ and $\{(i, 1, 1), 0 \leq i
\leq 2\}$, situated in the given elementary $3 \times 3$ square;
i.e., we can mark the point $(2, 2, 1)$ and shade the line $\{(i, 2,
1), 0 \leq i \leq 2\}$.

\looseness=1
Since the determinant of the matrix of values of the f\/ield at the
points of the elementary $3 \times 3$ square $\{(i, 0, 1), (i, 1,
1), (i, 2, 1), i = 0, 1, 2\}$ (on a middle normal section of our
cube) vanishes and, as it was noted above, the vectors of values of
the f\/ield $u$ at the points of the two shaded lines, $\{(0, j, 1), 0
\leq j \leq 2\}$ and $\{(1, j, 1), 0 \leq j \leq 2\}$, are linearly
independent, it follows immediately that the vector of values of the
f\/ield $u$ at the points of the line $\{(2, j, 1), 0 \leq j \leq 2\}$
is a linear combination of the vectors of values of the f\/ield $u$ at
the points of two other lines of this elementary $3 \times 3$
square, namely, two shaded lines of the same type, $\{(0, j, 1), 0
\leq j \leq 2\}$ and $\{(1, j, 1), 0 \leq j \leq 2\}$, situated in
the given elementary $3 \times 3$ square; i.e., we can shade the
line $\{(2, j, 1), 0 \leq j \leq 2\}$.

Let us consider one more elementary $3 \times 3$ square $\{(1, j,
0), (1, j, 1), (1, j, 2), j = 0, 1, 2\}$ in our cube (a middle
normal section of the cube). In this elementary square at the
present moment the values of the f\/ield $u$ are already determined at
eight points and the value of the f\/ield $u$ at the remaining ninth
point $(1, 2, 2)$ is determined by relation~\eqref{5}, i.e.,
by the condition that the determinant of the matrix of values of the
f\/ield at the points of this elementary $3 \times 3$ square vanishes,
because the corresponding $2 \times 2$ minor formed by initial data
is not equal to zero. Since, as it was noted above, the vectors of
values of the f\/ield $u$ at the points of the two shaded lines,
$\{(1, j, 0), 0 \leq j \leq 2\}$ and $\{(1, j, 1), 0 \leq j \leq
2\}$, are linearly independent, the vector of values of the f\/ield
$u$ at the points of the line $\{(1, j, 2), 0 \leq j \leq 2\}$ is a
linear combination of the vectors of values of the f\/ield $u$ at the
points of these two shaded lines of the same type, $\{(1, j, 0), 0
\leq j \leq 2\}$ and $\{(1, j, 1), 0 \leq j \leq 2\}$, situated in
the given elementary $3 \times 3$ square; i.e., we can mark the
point $(1, 2, 2)$ and shade the line $\{(1, j, 2), 0 \leq j \leq
2\}$.

Since the determinant of the matrix of values of the f\/ield at the
points of the elementary $3 \times 3$ square $\{(1, j, 0), (1, j,
1), (1, j, 2), j = 0, 1, 2\}$ (on a middle normal section of our
cube) vanishes and, as it was noted above, the vectors of values of
the f\/ield $u$ at the points of the two shaded lines, $\{(1, 0, k), 0
\leq k \leq 2\}$ and $\{(1, 1, k), 0 \leq k \leq 2\}$, are linearly
independent, it follows immediately that the vector of values of the
f\/ield $u$ at the points of the line $\{(1, 2, k), 0 \leq k \leq 2\}$
is a~linear combination of the vectors of values of the f\/ield $u$ at
the points of two other lines of this elementary $3 \times 3$
square, namely, two shaded lines of the same type, $\{(1, 0, k), 0
\leq k \leq 2\}$ and $\{(1, 1, k), 0 \leq k \leq 2\}$, situated in
the given elementary $3 \times 3$ square; i.e., we can shade the
line $\{(1, 2, k), 0 \leq k \leq 2\}$.

Let us consider one more elementary $3 \times 3$ square $\{(i, 1,
0), (i, 1, 1), (i, 1, 2), i = 0, 1, 2\}$ in our cube (a middle
normal section of the cube). In this elementary square at the
present moment the values of the f\/ield $u$ are already determined at
eight points and the value of the f\/ield $u$ at the remaining ninth
point $(2, 1, 2)$ is determined by relation~\eqref{5}, i.e.,
by the condition that the determinant of the matrix of values of the
f\/ield at the points of this elementary $3 \times 3$ square vanishes,
because the corresponding $2 \times 2$ minor formed by initial data
is not equal to zero. Since, as it was noted above, the vectors of
values of the f\/ield $u$ at the points of the two shaded lines,
$\{(i, 1, 0), 0 \leq i \leq 2\}$ and $\{(i, 1, 1), 0 \leq i \leq
2\}$, are linearly independent, the vector of values of the f\/ield
$u$ at the points of the line $\{(i, 1, 2), 0 \leq i \leq 2\}$ is a
linear combination of the vectors of values of the f\/ield $u$ at the
points of these two shaded lines of the same type, $\{(i, 1, 0), 0
\leq i \leq 2\}$ and $\{(i, 1, 1), 0 \leq i \leq 2\}$, situated in
the given elementary $3 \times 3$ square; i.e., we can mark the
point $(2, 1, 2)$ and shade the line $\{(i, 1, 2), 0 \leq i \leq
2\}$.

Since the determinant of the matrix of values of the f\/ield at the
points of the elementary $3 \times 3$ square $\{(i, 1, 0), (i, 1,
1), (i, 1, 2), i = 0, 1, 2\}$ (on a middle normal section of our
cube) vanishes and, as it was noted above, the vectors of values of
the f\/ield $u$ at the points of the two shaded lines, $\{(0, 1, k), 0
\leq k \leq 2\}$ and $\{(1, 1, k), 0 \leq k \leq 2\}$, are linearly
independent, it follows immediately that the vector of values of the
f\/ield $u$ at the points of the line $\{(2, 1, k), 0 \leq k \leq 2\}$
is a~linear combination of the vectors of values of the f\/ield $u$ at
the points of two other lines of this elementary $3 \times 3$
square, namely, two shaded lines of the same type, $\{(0, 1, k), 0
\leq k \leq 2\}$ and $\{(1, 1, k), 0 \leq k \leq 2\}$, situated in
the given elementary $3 \times 3$ square; i.e., we can shade the
line $\{(2, 1, k), 0 \leq k \leq 2\}$.

It remains to determine the value of the f\/ield $u$ only at one point
$(2, 2, 2)$ of our cube, and only three edges of the cube that
contain this point are still unshaded for the present.

In order to determine the value of the f\/ield $u$ at the remaining
lattice point $(2, 2, 2)$, it is necessary to show that in our cube
there is at least one elementary $3 \times 3$ square (bent or
unbent) containing this point $(2, 2, 2)$ and such that the
corresponding $2 \times 2$ minor in the $3 \times 3$ matrix of
values of the f\/ield $u$ at the points of this elementary $3 \times
3$ square is not equal to zero, since otherwise any value of the
f\/ield $u$ at the point $(2, 2, 2)$ could not be determined. Note
that in our cube the lattice point $(2, 2, 2)$ is the only one for
which in each elementary $3 \times 3$ square (bent or unbent)
containing this lattice point the corresponding $2 \times 2$ minor
is not completely formed by initial data.

Let us prove that the determinant of the $2 \times 2$ matrix of
values of the f\/ield $u$ at the points $\{(0, 1, 1), (1, 1, 1), (0,
2, 1), (1, 2, 1)\}$ of our cube is not equal to zero. Let us assume
that it va\-nishes. In this case, the $2$-vectors $(u_{0 1 1}, u_{0 2
1})$ and $(u_{1 1 1}, u_{1 2 1})$ must be linearly dependent, and
since the vector $(u_{0 1 1}, u_{0 2 1})$ of initial data must not
be zero (otherwise the corresponding $2 \times 2$ minor formed by
initial data must vanish), the vector $(u_{1 1 1}, u_{1 2 1})$ is
proportional to the vector $(u_{0 1 1}, u_{0 2 1})$: \begin{gather} (u_{1 1 1},
u_{1 2 1}) = \lambda (u_{0 1 1}, u_{0 2 1}). \label{6x} \end{gather} On the
other hand, as it was noted above, the vector $(u_{1 0 1}, u_{1 1
1}, u_{1 2 1})$ of values of the f\/ield $u$ is a~linear combination
of the basic vectors of the same type $(u_{0 0 0}, u_{0 1 0}, u_{0 2
0})$ and $(u_{0 0 1}, u_{0 1 1}, u_{0 2 1})$:
\begin{gather}
(u_{1 0 1}, u_{1 1 1}, u_{1 2 1}) = \alpha (u_{0 0 0}, u_{0 1
0}, u_{0 2 0}) + \beta (u_{0 0 1}, u_{0 1 1}, u_{0 2 1}),
\label{7x}
\end{gather}
and in this case \[ (u_{1 1 1}, u_{1 2 1}) = \alpha (u_{0 1 0},
u_{0 2 0}) + \beta (u_{0 1 1}, u_{0 2 1}). \] Using relation~\eqref{6x}
we obtain \[ \lambda (u_{0 1 1}, u_{0 2 1}) = \alpha
(u_{0 1 0}, u_{0 2 0}) + \beta (u_{0 1 1}, u_{0 2 1}). \] Hence,
since the $2$-vectors $(u_{0 1 1}, u_{0 2 1})$ and $(u_{0 1 0}, u_{0
2 0})$ are linearly independent (otherwise the corresponding $2
\times 2$ minor formed by initial data must vanish), we have $\alpha
= 0$, i.e., relation~\eqref{7x} assumes the form
\begin{gather*} (u_{1 0 1},
u_{1 1 1}, u_{1 2 1}) = \beta (u_{0 0 1}, u_{0 1 1}, u_{0 2 1}),
\end{gather*}
but this is impossible, because in this case the
corresponding $2 \times 2$ minor formed by initial data must vanish.

Thus, it is proved that the determinant of the $2 \times 2$ matrix
of values of the f\/ield $u$ at the points $\{(0, 1, 1), (1, 1, 1),
(0, 2, 1), (1, 2, 1)\}$ of our cube is not equal to zero.

Let us consider the bent elementary $3 \times 3$ square $\{(i, 1,
1), (i, 2, 1), (i, 2, 2), i = 0, 1, 2\}$ in our cube. In this
elementary square at the present moment the values of the f\/ield $u$
are already determined at eight points and the value of the f\/ield
$u$ at the remaining ninth point $(2, 2, 2)$ is determined by
relation~\eqref{5}, i.e., by the condition that the
determinant of the matrix of values of the f\/ield at the lattice
points of this elementary $3 \times 3$ square vanishes, because we
proved that the corresponding $2 \times 2$ minor is not equal to
zero. Since the vectors of values of the f\/ield $u$ at the points of
the two shaded lines, $\{(i, 2, 1), 0 \leq i \leq 2\}$ and $\{(i, 1,
1), 0 \leq i \leq 2\}$, are linearly independent (otherwise the
nonzero $2 \times 2$ minor considered above must vanish), the vector
of values of the f\/ield $u$ at the points of the line $\{(i, 2, 2), 0
\leq i \leq 2\}$ is a linear combination of the vectors of values of
the f\/ield $u$ at the points of these two shaded lines of the same
type, $\{(i, 2, 1), 0 \leq i \leq 2\}$ and $\{(i, 1, 1), 0 \leq i
\leq 2\}$, situated in the given bent elementary $3 \times 3$
square; i.e., we can mark the point $(2, 2, 2)$ and shade the line
$\{(i, 2, 2), 0 \leq i \leq 2\}$.

Now the values of the f\/ield $u$ are determined already at all points
of our cube, and it remains to shade two edges of the cube.

At f\/irst we prove that the determinant of the $2 \times 2$ matrix of
values of the f\/ield $u$ at the points $\{(0, 1, 1), (1, 1, 1), (0,
1, 2), (1, 1, 2)\}$ of our cube is not equal to zero. Let us assume
that it vanishes. In this case, the $2$-vectors $(u_{0 1 1}, u_{0 1
2})$ and $(u_{1 1 1}, u_{1 1 2})$ must be linearly dependent.
Moreover, both these vectors are nonzero, since otherwise the
corresponding $2 \times 2$ minor formed by initial data must vanish.
Indeed, let us assume, for example, that the $2$-vector $(u_{0 1 1},
u_{0 1 2})$ is zero, i.e., $u_{0 1 1} = 0, u_{0 1 2} = 0.$ Then, in
the bent elementary $3 \times 3$ square $\{(0, 1, k), (0, 0, k), (1,
0, k), k = 0, 1, 2\}$ in our cube, the determinant of the matrix of
values of the f\/ield $u$ at the points of this bent elementary $3
\times 3$ square is equal, except for sign, to the product of the
value $u_{0 1 0}$ of the f\/ield $u$ by the corresponding $2 \times 2$
minor formed by initial data, and this $2 \times 2$ minor is not
equal to zero by our condition on initial data. Thus, as the
determinant of the matrix of values of the f\/ield at the points of
this bent elementary $3 \times 3$ square vanishes, it follows that
$u_{0 1 0} = 0$, and since $u_{0 1 1} = 0$, the corresponding $2
\times 2$ minor formed by initial data is equal to zero, but this is
impossible. Similarly, it can be proved that the $2$-vector $(u_{1 1
1}, u_{1 1 2})$ is also nonzero. Indeed, let us assume that the
$2$-vector $(u_{1 1 1}, u_{1 1 2})$ is zero, i.e., $u_{1 1 1} = 0,
u_{1 1 2} = 0.$ Then, in the bent elementary $3 \times 3$ square
$\{(1, 1, k), (1, 0, k), (0, 0, k), k = 0, 1, 2\}$ in our cube, the
determinant of the matrix of values of the f\/ield $u$ at the points
of this bent elementary $3 \times 3$ square is equal, except for
sign, to the product of the value $u_{1 1 0}$ of the f\/ield $u$ by
the corresponding $2 \times 2$ minor formed by initial data, and
this $2 \times 2$ minor is not equal to zero by our condition on
initial data. Thus, as the determinant of the matrix of values of
the f\/ield at the points of this bent elementary $3 \times 3$ square
vanishes, it follows that $u_{1 1 0} = 0$, and since $u_{1 1 1} =
0$, the corresponding $2 \times 2$ minor formed by initial data is
equal to zero, but this is impossible. So we have proved that the
$2$-vectors $(u_{0 1 1}, u_{0 1 2})$ and $(u_{1 1 1}, u_{1 1 2})$
are nonzero. Moreover, these $2$-vectors must be linearly dependent
under our assumption. Therefore, each of these $2$-vectors is
proportional to the other: \begin{gather} (u_{0 1 1}, u_{0 1 2}) = \lambda
(u_{1 1 1}, u_{1 1 2}), \qquad (u_{1 1 1}, u_{1 1 2}) = \mu (u_{0 1
1}, u_{0 1 2}). \label{6xx} \end{gather}

On the other hand, since the determinant of the matrix of values of
the f\/ield at the points of the bent elementary $3 \times 3$ square
$\{(1, 0, k), (1, 1, k), (0, 1, k), k = 0, 1, 2\}$ vanishes and, as
it was noted above, the vectors of values of the f\/ield $u$ at the
points of the two shaded lines, $\{(1, 1, k), 0 \leq k \leq 2\}$ and
$\{(1, 0, k), 0 \leq k \leq 2\}$, are linearly independent, it
follows immediately that the vector of values of the f\/ield $u$ at
the points of the line $\{(0, 1, k), 0 \leq k \leq 2\}$ is a linear
combination of the vectors of values of the f\/ield $u$ at the points
of two other lines of this bent elementary $3 \times 3$ square,
namely, two shaded lines of the same type, $\{(1, 1, k), 0 \leq k
\leq 2\}$ and $\{(1, 0, k), 0 \leq k \leq 2\}$, situated in the
given bent elementary $3 \times 3$ square:
\begin{gather} (u_{0 1 0}, u_{0 1 1}, u_{0 1 2}) = \alpha (u_{1 1 0}, u_{1 1
1}, u_{1 1 2}) + \beta (u_{1 0 0}, u_{1 0 1}, u_{1 0 2}),
\label{7xx}
\end{gather}
and in this case \[ (u_{0 1 1}, u_{0 1 2}) = \alpha
(u_{1 1 1}, u_{1 1 2}) + \beta (u_{1 0 1}, u_{1 0 2}). \] Using
relation~\eqref{6xx}, we obtain \[ \lambda (u_{1 1 1}, u_{1 1 2}) =
\alpha (u_{1 1 1}, u_{1 1 2}) + \beta (u_{1 0 1}, u_{1 0 2}). \]

If the $2$-vectors $(u_{1 1 1}, u_{1 1 2})$ and $(u_{1 0 1}, u_{1 0
2})$ are linearly independent, i.e., the determinant of the matrix
of values of the f\/ield $u$ at the points $\{(1, 1, 1), (1, 0, 1),
(1, 0, 2), (1, 1, 2)\}$ of our cube is not equal to zero, then
$\beta = 0$ and relation~\eqref{7xx} assumes the form
\begin{gather*}
(u_{0 1 0}, u_{0 1 1}, u_{0 1 2}) = \alpha (u_{1 1 0}, u_{1 1 1}, u_{1 1
2}), 
\end{gather*}
but this is impossible, since in this case the
corresponding $2 \times 2$ minor formed by initial data must be
equal to zero.
Hence, under our assumptions the $2$-vectors $(u_{1 1 1}, u_{1 1
2})$ and $(u_{1 0 1}, u_{1 0 2})$ must be linearly dependent, i.e.,
the determinant of $2 \times 2$ matrix of values of the f\/ield $u$ at
the points $\{(1, 1, 1), (1, 0, 1), (1, 0, 2), (1, 1, 2)\}$ of our
cube vanishes. Since the $2$-vector $(u_{1 1 1}, u_{1 1 2})$ is
nonzero and the determinant of the $2 \times 2$ matrix of values of
the f\/ield $u$ at the points $\{(1, 1, 1), (1, 0, 1), (1, 0, 2), (1,
1, 2)\}$ of our cube vanishes, it follows that the $2$-vector $(u_{1
0 1}, u_{1 0 2})$ is proportional to the $2$-vector $(u_{1 1 1},
u_{1 1 2})$: \begin{gather} (u_{1 0 1}, u_{1 0 2}) = \nu (u_{1 1 1}, u_{1 1
2}). \label{9xx}
\end{gather}

Since the determinant of the matrix of values of the f\/ield at the
points of the bent elementary $3 \times 3$ square $\{(1, 1, k), (0,
1, k), (0, 0, k), k = 0, 1, 2\}$ vanishes and, as it was noted
above, the vectors of values of the f\/ield $u$ at the points of the
two shaded lines, $\{(0, 1, k), 0 \leq k \leq 2\}$ and $\{(0, 0, k),
0 \leq k \leq 2\}$, are linearly independent, it follows immediately
that the vector of values of the f\/ield $u$ at the points of the line
$\{(1, 1, k), 0 \leq k \leq 2\}$ is a linear combination of the
vectors of values of the f\/ield $u$ at the points of two other lines
of this bent elementary $3 \times 3$ square, namely, two shaded
lines of the same type, $\{(0, 1, k), 0 \leq k \leq 2\}$ and $\{(0,
0, k), 0 \leq k \leq 2\}$, situated in the given bent elementary $3
\times 3$ square:
\begin{gather} (u_{1 1 0}, u_{1 1 1}, u_{1 1 2}) = \alpha (u_{0 1 0}, u_{0 1
1}, u_{0 1 2}) + \beta (u_{0 0 0}, u_{0 0 1}, u_{0 0 2}),
\label{7xxz}
\end{gather} and in this case \[ (u_{1 1 1}, u_{1 1 2}) =
\alpha (u_{0 1 1}, u_{0 1 2}) + \beta (u_{0 0 1}, u_{0 0 2}). \]
Using relation~\eqref{6xx}, we obtain \[ \mu (u_{0 1 1}, u_{0 1 2})
= \alpha (u_{0 1 1}, u_{0 1 2}) + \beta (u_{0 0 1}, u_{0 0 2}). \]

If the $2$-vectors $(u_{0 1 1}, u_{0 1 2})$ and $(u_{0 0 1}, u_{0 0
2})$ are linearly independent, i.e., the determinant of the matrix
of values of the f\/ield $u$ at the points $\{(0, 1, 1), (0, 0, 1),
(0, 0, 2), (0, 1, 2)\}$ of our cube is not equal to zero, then
$\beta = 0$ and relation~\eqref{7xxz} assumes the form
\begin{gather*} (u_{1 1
0}, u_{1 1 1}, u_{1 1 2}) = \alpha (u_{0 1 0}, u_{0 1 1}, u_{0 1
2}), 
\end{gather*} but this is impossible, since in this case the
corresponding $2 \times 2$ minor formed by initial data must be
equal to zero.
Hence, under our assumptions the $2$-vectors $(u_{0 1 1}, u_{0 1
2})$ and $(u_{0 0 1}, u_{0 0 2})$ must be linearly dependent, i.e.,
the determinant of the $2 \times 2$ matrix of values of the f\/ield~$u$ at the points $\{(0, 1, 1), (0, 0, 1), (0, 0, 2), (0, 1, 2)\}$
of our cube vanishes. Since the $2$-vector $(u_{0 1 1}, u_{0 1 2})$
is nonzero and the determinant of the $2 \times 2$ matrix of values
of the f\/ield $u$ at the points $\{(0, 1, 1), (0, 0, 1), (0, 0, 2),
(0, 1, 2)\}$ of our cube vanishes, it follows that the $2$-vector
$(u_{0 0 1}, u_{0 0 2})$ is proportional to the $2$-vector $(u_{0 1
1}, u_{0 1 2})$: \[ (u_{0 0 1}, u_{0 0 2}) = \varkappa \, (u_{0 1
1}, u_{0 1 2}). \]

Using relation~\eqref{6xx}, we obtain \[ (u_{0 0 1}, u_{0 0 2}) =
\varkappa \, (u_{0 1 1}, u_{0 1 2}) = \varkappa \, \lambda (u_{1 1
1}, u_{1 1 2}), \] and from \eqref{9xx} we have
\begin{gather*} (u_{1 0 1},
u_{1 0 2}) = \nu (u_{1 1 1}, u_{1 1 2}), 
\end{gather*} i.e., the
$2$-vectors $(u_{0 0 1}, u_{0 0 2})$ and $(u_{1 0 1}, u_{1 0 2})$
are linearly dependent, but this is impossible, since in this case
the corresponding $2 \times 2$ minor formed by initial data must be
equal to zero.

Thus, it is proved that the determinant of the $2 \times 2$ matrix
of values of the f\/ield $u$ at the points $\{(0, 1, 1), (1, 1, 1),
(0, 1, 2), (1, 1, 2)\}$ of our cube is not equal to zero.

Since the determinant of the matrix of values of the f\/ield $u$ at
the points of the bent elementary $3 \times 3$ square $\{(i, 0, 2),
(i, 1, 2), (i, 1, 1), i = 0, 1, 2\}$ vanishes (the three lines
$\{(i, 0, 2), i = 0, 1, 2\}$, $\{(i, 1, 2), i = 0, 1, 2\}$ and
$\{(i, 1, 1), i = 0, 1, 2\}$ are shaded in this bent elementary $3
\times 3$ square), it follows immediately that the three vectors of
values of the f\/ield $u$ at the points of the bent lines $\{(0, 0,
2), (0, 1, 2), (0, 1, 1)\}$, $\{(1, 0, 2), (1, 1, 2), (1, 1, 1)\}$
and $\{(2, 0, 2), (2, 1, 2), (2, 1, 1)\}$ are linearly dependent;
moreover, the vectors of values of the f\/ield $u$ at the points of
the bent lines $\{(0, 0, 2), (0, 1, 2), (0, 1, 1)\}$ and $\{(1, 0,
2), (1, 1, 2), (1, 1, 1)\}$ are linearly independent, since
otherwise the nonzero determinant of the $2 \times 2$ matrix of
values of the f\/ield $u$ at the points $\{(0, 1, 1), (1, 1, 1), (0,
1, 2), (1, 1, 2)\}$ of our cube must vanish. Thus, the vector of
values of the f\/ield $u$ at the points of the bent line $\{(2, 0, 2),
(2, 1, 2), (2, 1, 1)\}$ is a linear combination of the vectors of
values of the f\/ield $u$ at the points of the bent lines $\{(0, 0,
2), (0, 1, 2), (0, 1, 1)\}$ and $\{(1, 0, 2), (1, 1, 2), (1, 1,
1)\}$: \begin{gather} (u_{2 0 2}, u_{2 1 2}, u_{2 1 1}) = \alpha (u_{0 0 2},
u_{0 1 2}, u_{0 1 1}) + \beta (u_{1 0 2}, u_{1 1 2}, u_{1 1 1}).
\label{7xxy} \end{gather}

Similarly, since the determinant of the matrix of values of the
f\/ield at the points of the bent ele\-mentary $3 \times 3$ square
$\{(i, 2, 2), (i, 1, 2), (i, 1, 1), i = 0, 1, 2\}$ vanishes (the
three lines $\{(i, 2, 2), i = 0, 1, 2\}$, $\{(i, 1, 2), i = 0, 1,
2\}$ and $\{(i, 1, 1), i = 0, 1, 2\}$ are shaded in this bent
elementary $3 \times 3$ square), it follows immediately that the
vectors of values of the f\/ield $u$ at the points of the bent lines
$\{(0, 2, 2), (0, 1, 2), (0, 1, 1)\}$, $\{(1, 2, 2), (1, 1, 2), (1,
1, 1)\}$ and $\{(2, 2, 2), (2, 1, 2), (2, 1, 1)\}$ are linearly
dependent; moreover, the vectors of values of the f\/ield $u$ at the
points of the bent lines $\{(0, 2, 2), (0, 1, 2), (0, 1, 1)\}$ and
$\{(1, 2, 2), (1, 1, 2), (1, 1, 1)\}$ are linearly independent,
since otherwise the nonzero determinant of the $2 \times 2$ matrix
of values of the f\/ield $u$ at the points $\{(0, 1, 1), (1, 1, 1),
(0, 1, 2), (1, 1, 2)\}$ of our cube must vanish. Thus, in the bent
elementary $3 \times 3$ square under consideration, the vector of
values of the f\/ield $u$ at the points of the bent line $\{(2, 2, 2),
(2, 1, 2), (2, 1, 1)\}$ is a linear combination of the vectors of
values of the f\/ield $u$ at the points of the bent lines $\{(0, 2,
2), (0, 1, 2), (0, 1, 1)\}$ and $\{(1, 2, 2), (1, 1, 2), (1, 1,
1)\}$:
\begin{gather}
(u_{2 2 2}, u_{2 1 2}, u_{2 1 1}) = \gamma (u_{0 2 2},
u_{0 1 2}, u_{0 1 1}) + \delta (u_{1 2 2}, u_{1 1 2}, u_{1 1 1}).
\label{7xxyz}
\end{gather} From relations~\eqref{7xxy} and \eqref{7xxyz}, we
obtain respectively
\begin{gather*} (u_{2 1 2}, u_{2 1 1}) = \alpha (u_{0 1 2},
u_{0 1 1}) + \beta (u_{1 1 2}, u_{1 1 1}) 
\end{gather*} and
\begin{gather*}
(u_{2 1 2}, u_{2 1 1}) = \gamma (u_{0 1 2}, u_{0 1 1}) + \delta
(u_{1 1 2}, u_{1 1 1}), 
\end{gather*} whence it follows
immediately that $\alpha = \gamma$ and $\beta = \delta$, since the
determinant of the $2 \times 2$ matrix of values of the f\/ield $u$ at
the points $\{(0, 1, 1), (1, 1, 1), (0, 1, 2), (1, 1, 2)\}$ of our
cube is not equal to zero and the $2$-vectors $(u_{0 1 1}, u_{0 1
2})$ and $(u_{1 1 1}, u_{1 1 2})$ are linearly independent. From
relations~\eqref{7xxy} and \eqref{7xxyz}, we obtain respectively \begin{gather}
(u_{2 0 2}, u_{2 1 2}) = \alpha (u_{0 0 2}, u_{0 1 2}) + \beta (u_{1
0 2}, u_{1 1 2}) \label{7xxyb} \end{gather} and
\begin{gather} (u_{2 2 2}, u_{2 1 2}) =
\gamma (u_{0 2 2}, u_{0 1 2}) + \delta (u_{1 2 2}, u_{1 1 2}).
\label{7xxyzb}
\end{gather}
Since $\alpha = \gamma$ and $\beta = \delta$,
from relations~\eqref{7xxyb} and \eqref{7xxyzb}
\begin{gather*} (u_{2 0 2}, u_{2
1 2}, u_{2 2 2}) = \alpha (u_{0 0 2}, u_{0 1 2}, u_{0 2 2}) + \beta
(u_{1 0 2}, u_{1 1 2}, u_{1 2 2}), 
\end{gather*} i.e., the
vector of values of the f\/ield $u$ at the points of the line $\{(2,
2, 2), (2, 1, 2), (2, 0, 2)\}$ is a~linear combination of the
vectors of values of the f\/ield $u$ at the points of the two shaded
lines $\{(0, 2, 2), (0, 1, 2), (0, 0, 2)\}$ and $\{(1, 2, 2), (1, 1,
2), (1, 0, 2)\}$, and hence we can shade also the line $\{(2, 2, 2),
(2, 1, 2), (2, 0, 2)\}$ in our cube.

Let us prove now that the determinant of the $2 \times 2$ matrix of
values of the f\/ield $u$ at the points $\{(1, 0, 1), (1, 1, 1), (2,
1, 1), (2, 0, 1)\}$ of our cube is not equal to zero. Let us assume
that it vanishes. In this case, the $2$-vectors $(u_{1 1 1}, u_{2 1
1})$ and $(u_{1 0 1}, u_{2 0 1})$ must be linearly dependent.
Moreover, the vector $(u_{1 0 1}, u_{2 0 1})$ is nonzero, since
otherwise the corresponding $2 \times 2$ minor formed by initial
data must vanish. In this case the $2$-vector $(u_{1 1 1}, u_{2 1
1})$ must be proportional to the $2$-vector $(u_{1 0 1}, u_{2 0
1})$: \begin{gather} (u_{1 1 1}, u_{2 1 1}) = \lambda (u_{1 0 1}, u_{2 0 1}).
\label{6xxb} \end{gather}

On the other hand, since the determinant of the matrix of values of
the f\/ield at the points of the bent elementary $3 \times 3$ square
$\{(i, 0, 0), (i, 0, 1), (i, 1, 1), i = 0, 1, 2\}$ vanishes and the
vectors of values of the f\/ield $u$ at the points of the two basic
lines $\{(i, 0, 0), 0 \leq i \leq 2\}$ and $\{(i, 0, 1), 0 \leq i
\leq 2\}$ are linearly independent, it follows immediately that the
vector of values of the f\/ield $u$ at the points of the line $\{(i,
1, 1), 0 \leq i \leq 2\}$ is a linear combination of the vectors of
values of the f\/ield $u$ at the points of the two lines of this bent
elementary $3 \times 3$ square, namely, the two basic lines of the
same type $\{(i, 0, 0), 0 \leq i \leq 2\}$ and $\{(i, 0, 1), 0 \leq
i \leq 2\}$, situated in the given bent elementary $3 \times 3$
square:
\begin{gather} (u_{0 1 1}, u_{1 1 1}, u_{2 1 1}) = \alpha (u_{0 0 1}, u_{1 0
1}, u_{2 0 1}) + \beta (u_{0 0 0}, u_{1 0 0}, u_{2 0 0}),
\label{7xxb}
\end{gather}
and in this case \[ (u_{1 1 1}, u_{2 1 1}) =
\alpha (u_{1 0 1}, u_{2 0 1}) + \beta (u_{1 0 0}, u_{2 0 0}). \]
Using relation~\eqref{6xxb}, we obtain \[ \lambda (u_{1 0 1}, u_{2
0 1}) = \alpha (u_{1 0 1}, u_{2 0 1}) + \beta (u_{1 0 0}, u_{2 0
0}). \] Since the $2$-vectors $(u_{1 0 1}, u_{2 0 1})$ and $(u_{1 0
0}, u_{2 0 0})$ are linearly independent (otherwise the
corresponding $2 \times 2$ minor formed by initial data must
vanish), we have $\beta = 0$ and relation~\eqref{7xxb} assumes the
form \begin{gather*} (u_{0 1 1}, u_{1 1 1}, u_{2 1 1}) = \alpha (u_{0 0 1}, u_{1
0 1}, u_{2 0 1}), 
\end{gather*} but this is impossible, because
in this case the corresponding $2 \times 2$ minor formed by initial
data must vanish.

Thus, it is proved that the determinant of the $2 \times 2$ matrix
of values of the f\/ield $u$ at the points $\{(1, 0, 1), (1, 1, 1),
(2, 1, 1), (2, 0, 1)\}$ of our cube is not equal to zero.

Since the determinant of the matrix of values of the f\/ield at the
points of the bent elementary $3 \times 3$ square $\{(2, j, 0), (2,
j, 1), (1, j, 1), j = 0, 1, 2\}$ vanishes (the three lines $\{(2, j,
0), j = 0, 1, 2\}$, $\{(2, j, 1), j = 0, 1, 2\}$ and $\{(1, j, 1), j
= 0, 1, 2\}$ are shaded in this bent elementary $3 \times 3$
square), it follows immediately that the vectors of values of the
f\/ield $u$ at the points of the bent lines $\{(2, 0, 0), (2, 0, 1),
(1, 0, 1)\}$, $\{(2, 1, 0), (2, 1, 1), (1, 1, 1)\}$ and $\{(2, 2,
0), (2, 2, 1), (1, 2, 1)\}$ are linearly dependent; moreover, the
vectors of values of the f\/ield $u$ at the points of the bent lines
$\{(2, 0, 0), (2, 0, 1), (1, 0, 1)\}$ and $\{(2, 1, 0), (2, 1, 1),
(1, 1, 1)\}$ are linearly independent, since otherwise the nonzero
determinant of the $2 \times 2$ matrix of values of the f\/ield $u$ at
the points $\{(1, 0, 1), (1, 1, 1), (2, 1, 1), (2, 0, 1)\}$ of our
cube must vanish. Thus, the vector of values of the f\/ield $u$ at the
points of the bent line $\{(2, 2, 0), (2, 2, 1), (1, 2, 1)\}$ is a
linear combination of the vectors of values of the f\/ield $u$ at the
points of the bent lines $\{(2, 0, 0), (2, 0, 1), (1, 0, 1)\}$ and
$\{(2, 1, 0), (2, 1, 1), (1, 1, 1)\}$: \begin{gather} (u_{2 2 0}, u_{2 2 1},
u_{1 2 1}) = \alpha (u_{2 0 0}, u_{2 0 1}, u_{1 0 1}) + \beta (u_{2
1 0}, u_{2 1 1}, u_{1 1 1}). \label{7xxyb2} \end{gather}

Similarly, since the determinant of the matrix of values of the
f\/ield at the points of the bent elementary $3 \times 3$ square
$\{(2, j, 2), (2, j, 1), (1, j, 1), j = 0, 1, 2\}$ vanishes (the
three lines $\{(2, j, 2), j = 0, 1, 2\}$, $\{(2, j, 1), j = 0, 1,
2\}$ and $\{(1, j, 1), j = 0, 1, 2\}$ are shaded in this bent
elementary $3 \times 3$ square), it follows immediately that the
vectors of values of the f\/ield $u$ at the points of the bent lines
$\{(1, 0, 1), (2, 0, 1), (2, 0, 2)\}$, $\{(1, 1, 1), (2, 1, 1), (2,
1, 2)\}$ and $\{(1, 2, 1), (2, 2, 1), (2, 2, 2)\}$ are linearly
dependent; moreover, the vectors of values of the f\/ield $u$ at the
points of the bent lines $\{(1, 0, 1), (2, 0, 1), (2, 0, 2)\}$ and
$\{(1, 1, 1), (2, 1, 1), (2, 1, 2)\}$ are linearly independent,
since otherwise the nonzero determinant of the $2 \times 2$ matrix
of values of the f\/ield $u$ at the points $\{(1, 0, 1), (1, 1, 1),
(2, 1, 1), (2, 0, 1)\}$ of our cube must vanish. Thus, in the bent
elementary $3 \times 3$ square under consideration, the vector of
values of the f\/ield $u$ at the points of the bent line $\{(1, 2, 1),
(2, 2, 1), (2, 2, 2)\}$ is a linear combination of the vectors of
values of the f\/ield $u$ at the points of the bent lines $\{(1, 0,
1), (2, 0, 1), (2, 0, 2)\}$ and $\{(1, 1, 1), (2, 1, 1), (2, 1,
2)\}$:  \begin{gather} (u_{2 2 2}, u_{2 2 1}, u_{1 2 1}) = \gamma (u_{2 0 2},
u_{2 0 1}, u_{1 0 1}) + \delta (u_{2 1 2}, u_{2 1 1}, u_{1 1 1}).
\label{7xxyzb2} \end{gather} From relations~\eqref{7xxyb2} and
\eqref{7xxyzb2}, we obtain respectively
\begin{gather*} (u_{2 2 1}, u_{1 2 1}) =
\alpha (u_{2 0 1}, u_{1 0 1}) + \beta (u_{2 1 1}, u_{1 1 1})
\end{gather*} and
\begin{gather*}
(u_{2 2 1}, u_{1 2 1}) = \gamma (u_{2 0
1}, u_{1 0 1}) + \delta (u_{2 1 1}, u_{1 1 1}),
\end{gather*}
whence it follows immediately that $\alpha = \gamma$ and $\beta =
\delta$, since the determinant of the $2 \times 2$ matrix of values
of the f\/ield $u$ at the points $\{(1, 0, 1), (1, 1, 1), (2, 1, 1),
(2, 0, 1)\}$ of our cube is not equal to zero and the $2$-vectors
$(u_{2 0 1}, u_{1 0 1})$ and $(u_{2 1 1}, u_{1 1 1})$ are linearly
independent. From relations~\eqref{7xxyb2} and \eqref{7xxyzb2}, we
obtain respectively
\begin{gather} (u_{2 2 0}, u_{2 2 1}) = \alpha (u_{2 0 0},
u_{2 0 1}) + \beta (u_{2 1 0}, u_{2 1 1}) \label{7xxybb2}
\end{gather} and
\begin{gather}
(u_{2 2 2}, u_{2 2 1}) = \gamma (u_{2 0 2}, u_{2 0 1}) + \delta
(u_{2 1 2}, u_{2 1 1}). \label{7xxyzbb2}
\end{gather} Since $\alpha = \gamma$
and $\beta = \delta$, from relations~\eqref{7xxybb2} and
\eqref{7xxyzbb2}
\begin{gather*} (u_{2 2 2}, u_{2 2 1}, u_{2 2 0}) = \alpha
(u_{2 0 2}, u_{2 0 1}, u_{2 0 0}) + \beta (u_{2 1 2}, u_{2 1 1},
u_{2 1 0}), 
\end{gather*} i.e., the vector of values of the
f\/ield $u$ at the points of the line $\{(2, 2, 2), (2, 2, 1), (2, 2,
0)\}$ is a~linear combination of the vectors of values of the f\/ield
$u$ at the points of the two shaded lines $\{(2, 0, 2), (2, 0, 1),
(2, 0, 0)\}$ and $\{(2, 1, 2), (2, 1, 1), (2, 1, 0)\}$, and hence we
can shade also the line $\{(2, 2, 2), (2, 2, 1), (2, 2, 0)\}$ in our
cube.

Thus, the values of the f\/ield $u$ are determined at all points of
our cube, and all lines of the cube are shaded now. The theorem is
proved.
\end{proof}

Moreover, we have proved a {\it considerably stronger principle of
consistency on the cubic lattice for determinants}.

\begin{theorem}[\cite{2}]\label{theorem2}
For arbitrary generic initial data, the
nonlinear discrete equation~\eqref{5} can be satisfied in a
consistent way and simultaneously on each set of points of three
lines $P_l$, $1 \leq l \leq 3$, of the cubic lattice $\mathbb{Z}^3$
of the form $P_l = \{(i, r_l, s_l), a \leq i \leq a + 2\}$, $1 \leq
l \leq 3$, where~$a$,~$r_l$, and~$s_l$, $1 \leq l \leq 3$, are
arbitrary fixed integers {\rm (}$x$-type lines{\rm )}, as well as on
each set of points of three lines $Q_l$, $1 \leq l \leq 3$, of the
cubic lattice $\mathbb{Z}^3$ of the form $Q_l = \{(r_l, j, s_l), a
\leq j \leq a + 2\}$, $1 \leq l \leq 3$, where $a$, $r_l$, and
$s_l$, $1 \leq l \leq 3$, are arbitrary fixed integers {\rm
(}$y$-type lines{\rm )}, and on each set of points of three lines
$R_l$, $1 \leq l \leq 3$, of the cubic lattice $\mathbb{Z}^3$ of the
form $R_l = \{(r_l, s_l, k), a \leq k \leq a + 2\}$, $1 \leq l \leq
3$, where $a$, $r_l$, and $s_l$, $1 \leq l \leq 3$, are arbitrary
fixed integers {\rm (}$z$-type lines{\rm )}. Moreover, in this case
the discrete equation~\eqref{5} will be satisfied in a
consistent way and simultaneously on each set of points of special
form lying on three bent lines $S_l$, $1 \leq l \leq 3$, of the same
type in the cubic lattice $\mathbb{Z}^3$, for example, of the form
$S_l = \{(a, r_l, s), (a + 1, r_l, s), (a + 1, r_l, s + 1)\}$, $1
\leq l \leq 3$, where $a$, $r_l$, and $s$, $1 \leq l \leq 3$, are
arbitrary fixed integers, of the form $S_l = \{(a, s, r_l), (a + 1,
s, r_l), (a + 1, s + 1, r_l)\}$, $1 \leq l \leq 3$, where $a$,
$r_l$, and $s$, $1 \leq l \leq 3$, are arbitrary fixed integers, or
of the form $S_l = \{(r_l, s, a + 1), (r_l, s, a), (r_l, s + 1,
a)\}$, $1 \leq l \leq 3$, where $a$, $r_l$, and $s$, $1 \leq l \leq
3$, are arbitrary fixed integers.
\end{theorem}

The following {\it principle of consistency on the cubic lattice for
determinants} also holds.

Let us consider an arbitrary line $P$ (bent or unbent) given by
three arbitrary neighboring points in the cubic lattice~$\mathbb{Z}^3$. We consider an arbitrary set of three lines of the
cubic lattice~$\mathbb{Z}^3$ that are obtained from the line $P$ by
translations in the lattice by vectors parallel to the
(one-dimensional or two-dimensional) space orthogonal to the line~$P$ (i.e., orthogonal to the plane or to the straight line of~$P$
depending on whether the line $P$ is bent or unbent). Then, for
arbitrary generic initial data, the nonlinear discrete equation~\eqref{5}
can be satisf\/ied in a consistent way and simultaneously
on each such set of three lines of the cubic lattice~$\mathbb{Z}^3$.

Similar properties of consistency on cubic lattices hold for
determinants of arbitrary order $N \geq 2$ (see~\cite{1, 2}).

\subsection*{Acknowledgements}

The work was carried out under partial
f\/inancial support from the Russian Foundation for Basic Research
(project no.~09-01-00762) and from the programme ``Leading
Scientif\/ic Schools'' (project no. NSh-5413.2010.1).

\pdfbookmark[1]{References}{ref}
\LastPageEnding

\end{document}